# Efficient 1 GHz Ti:sapphire laser with improved broadband continuum in the infrared

Giovana T. Nogueira[1] and Flavio C. Cruz[1,2]

[1] *Instituto de Física Gleb Wataghin, Universidade Estadual de Campinas, CP.6165, Campinas, SP, 13083-970, Brazil*
[2] *JILA, National Institute of Standards and Technology and University of Colorado Department of Physics, University of Colorado, Boulder, CO 80309-0440*

We demonstrate a 1 GHz prismless femtosecond Ti:sapphire ring laser which emits 890 mW for 7.6W of pump power over a continuum extending from 585 to 1200 nm at -20 dB below the maximum. A broadband continuum is obtained without careful mirror dispersion compensation, with the net cavity group-delay-dispersion having -50 to +100 $fs^2$ oscillations from 700 to 900 nm. Further broadening is obtained by use of a slightly convex cavity mirror that increases self-phase modulation. 17% (75%) of the intracavity (output) power is generated in single-pass through the crystal, outside the cavity bandwidth and concentrated in the low gain infrared region from 960 to 1200 nm. This laser seems well suited for optical frequency metrology, possibly allowing easier stabilization of the carrier-to-envelope offset frequency without use of photonic fibers.

Femtosecond lasers are the basis of optical frequency combs, which have revolutionized frequency metrology and precision measurements in the recent past years[1]. They have been used, for example, for direct measurements of frequencies of several hundred THz[2], in optical atomic clocks[3] and for phase-sensitive nonlinear optics experiments[4]. For measurement and control of the carrier-to-envelope offset frequency[1] ($f_{ceo}$) in optical frequency combs, a femtosecond laser whose spectrum covers one octave is desirable. Although the use of a microstructure fiber offers a solution to this, there are problems for continuous and long-term operation, which is important for clocks and other applications. In this way, the advent of femtosecond lasers (in particular Ti:sapphire) emitting a broadband continuum[5,6,7,8,9] has attracted interest. These lasers have been demonstrated with stationary-wave, longer cavities operating with prisms in order to compensate for group delay dispersion (GDD), or with short, traveling-wave, prismless ring cavities that use chirped mirrors. In the first case, repetition rates are typically smaller, near 100 MHz or below, while the second case allows for repetition rates of one Gigahertz or higher, which are often preferred for optical frequency measurements. Measurement and stabilization of $f_{ceo}$ has also been demonstrated with such lasers[7,8,9]. In this letter, we report what to our knowledge is the most broadband femtosecond spectrum to date, obtained from a Ti:sapphire laser whose spectrum covers from 585 to 1200 nm at a level of -20 dB below the maximum. It operates stably with an output power of 890 mW for 7.6 W of pump power, and a repetition rate near 1 GHz. We discuss its characterization and the points which we found to be important for stable ultra broadband operation.

The main laser cavity described here consists of a 4-mirror bow-tie ring cavity comprising two curved mirrors (including the input coupler (IC)), with broadband high reflecting (HR) coatings and radius of curvature of 3 cm, a HR flat mirror and a flat output coupler (OC) with a transmission of 2 % from 730 to 870 nm. All mirrors are commercially available[10] and chirped, with GDD of -60 $fs^2$ from 720 to 990 nm, excluding the OC. In another cavity configuration the HR flat mirror has been replaced by a broadband HR convex mirror (R = 1 m) in order to enhance self-phase modulation (SPM)[5]. A 3 mm-long Brewster-cut Ti:sapphire crystal is placed at the cavity tighter waist between the curved mirrors. The $Ti^{3+}$ concentration is such that 70% of the power from a single-frequency pump laser at 532 nm is absorbed. The net cavity GDD is shown in Fig.1. Curve (a) is an average of four traces obtained by white-light interferometry[11], including the two curved mirrors, the convex mirror, the Ti:sapphire crystal, but not the OC. Curve (b) is the calculated GDD based on the curves of the same mirrors provided by the manufacturer, and combined with a Ti:sapphire crystal flat GDD of +174 $fs^2$ [12]. Curve (c) is the same as (b), but with the convex mirror replaced by the HR flat mirror. Curve (d) is the calculated GDD of the OC only, also provided by the manufacturer[10]. A few comments can be made from Fig.1: 1) the positive GDD of the crystal is nearly compensated by the negative GDD of the chirped mirrors, giving small negative net GDD in certain wavelength regions. Small negative net GDD is known to give short pulses[13, 14]; 2) oscillation compensation has not been important for the generation of a broadband spectrum; 3) the net-GDD is even positive near the Ti:sapphire gain peak, from 760 to 800 nm.

In addition to the cavity described above, we have also built two ring cavities with six mirrors and reduced repetition rate of 920 MHz. These cavities differ from the former one by the addition of two HR flat chirped mirrors (-70$fs^2$ from 720 to 1000 nm, oscillation compensated[10]) and by a different chirped IC (-40 $fs^2$ from 700 to 850 nm, HR from 650 to 900 nm). They differ from each other by the use of a flat HR chirped mirror (-60 $fs^2$) or an unchirped mirror, such that the net cavity GDD (mirrors and crystal) amounted to -120 $fs^2$ and -60 $fs^2$ respectively. The convex mirror was not used. In either case we could not generate a broadband continuum. Curve a in Figure 2 corresponds to the laser spectrum for either of these 6-mirror cavities, showing a Gaussian profile centered at 777 nm, with a full-width at half-maximum of 10 nm. Curve b corresponds to the spectrum for the 4-mirror cavity using the flat HR mirror instead of the convex one. In curves c and d the convex mirror has been used. Curves c and d differ from each other only by changes in the cavity alignment, as we explain below.

A significant difference between the 6-mirror cavities with excess negative net GDD (curve a in Fig.2) and the 4-mirror one (curves b-d of Fig.2), is their sensitivity for alignment and operation. We found that mode-locking was more difficult to achieve for the 6-mirror cavities, requiring careful adjustment of the curved mirror angles. Precise positioning of the curved mirrors with respect to the crystal was also required to eliminate a continuous-wave (cw) component. On the other hand the 4-mirror cavity mode-locks easily, with considerably less critical adjustments. No instabilities associated with possible Q-switching operation were observed, as can be seen from the inset in Fig.3 which shows a beat signal at the laser repetition rate. For this 4-mirror cavity using the convex mirror and for input powers above 6 Watts, as the curved mirror opposing the IC is moved towards the crystal, three distinct regions can be identified near the inner edge of the 2.6-mm long stability range for the distance between these curved mirrors. In the first region the laser oscillates mode-locked in one direction of the ring cavity. An intermediate region is more sensitive to alignment, with mode-locked oscillation alternating between both directions of the ring cavity, without any cw component. The third region is similar to the first, but the laser oscillates in the other direction with respect to the first region. We note that there are similarities between this description and the one given in ref. 5. Curve b) and c) in Fig. 2 are typical spectra for the laser operating either in the first or third regions. These spectra are quite similar to those in refs. 5, 9 and are obtained either with the flat or convex HR mirror, showing that a broadband spectrum is primarily obtained when the net cavity GDD is small. Curve d) in Fig. 2 corresponds to operation in the second region using the convex mirror, which adds further broadening. We have not determined if these three regions are specifically associated with use of the convex mirror, and for input powers below 6 Watts only the first and third regions are observed.

The spectrum of curve d of Fig. 2 is quite remarkable, extending from 585 to 1200 nm at the -20 dB level from the maximum at 986 nm, or at -10 dB with respect to the power at 800 nm. Since in our measurements the laser oscillation direction has always been opposed to the pump direction, the light that leaves the crystal reflects in the IC and is transmitted through the OC. The peaks in the laser spectrum at 660, 680 nm and 986 nm in curves c) and d) in Fig.2 occur at maxima in the OC transmission spectrum and the dips at 1190 and 1250 nm are simply due to minima of the IC reflection spectrum. Assuming that light outside the cavity bandwidth is generated in only a single-pass through the crystal, mainly by SPM and also by the Ti:sapphire gain, then a considerable increase in output power would be possible simply if the transmission/reflection of the IC/OC could be optimized. If the spectrum of curve d in Fig. 2 is corrected by the reflectivity of the IC and OC (Fig. 2), we obtain the intracavity power spectrum. By defining a cavity bandwidth (BW) from 685 to 960 nm, corresponding to points where the transmission of the OC increases to 20% (vertical lines in Fig.2), we integrated the intracavity power spectrum to compute which fraction is generated inside the cavity BW and which fraction is generated outside it, in single-pass through the crystal. We thus obtain the impressive number that 17% of the intracavity power is generated in single-pass mainly by SPM and also gain in the crystal. It follows that this fraction increases to 75% for the available power outside the cavity. Most of this power is concentrated in the infrared portion of the spectrum, from 960 to 1200 nm. It is possible that the broadband chirped mirrors, with HR coatings and negative GDD extending up to 990 nm, the tight focus in a relatively long crystal (soft-aperture KLM), and high-order dispersion all play important roles for this improved generation in the infrared.

We have also replaced the OC by another mirror with transmission reduced to 1% over the same spectral extension. In this case, the output power has been nearly the same, and the spectrum slightly narrower. The use of a microstructure fiber, previously employed to generate a broadband continuum from the 6-mirror cavity lasers, does not cause any further spectral broadening.

Figure 3 shows power curves for the 4-mirror cavity. The curve with triangles is for CW operation, with the cavity optimally aligned in the center of its stability curve. The high slope efficiency of 32 % (1.5 W for 5W pumping) and the low threshold (350 mW) are consequences of the compact design with tight waist size in the crystal[15]. The curve with circles corresponds to the spectrum of curve c in Fig. 2, obtained in the first region discussed above. We obtain a slope efficiency of 7 % and a threshold near 3 Watts for mode-locked operation. 330 mW (560 mW) is obtained for 5 W (7 W) of pump power. Although in this region cw oscillation can occur for pump powers lower than 3 W, it has not been observed above that pump power. Curves a, b and c in Fig. 2 were obtained for 5 W pumping and curve d was obtained for 8 W. However there was no significant variation of the output spectrum as a function of input power. The curve with squares in Fig. 3 corresponds to the broadest spectrum of curve d in Fig. 2, obtained in the second region discussed above. Here the curve is very nonlinear. It shows a bi-stable regime with hysteresis, with the power sometimes changing discontinuously in a way that depends on whether the pump power is increased or decreased. By only changing the pump power, we can also observe two output powers for the same pump power. Mode-locked in this second region has always led to a broad continuum. For the 6-mirror cavities we obtained output powers of 550 mW for 5 W of pump power.

In summary, we have demonstrated a 1 GHz fs Ti:sapphire prismless laser that emits a broadband continuum with an improved infrared spectrum. 75% of the output power is generated in a single-pass through the crystal, mainly by self-phase modulation, and is concentrated in the spectral region from 960 to 1200 nm, outside the laser cavity bandwidth and where the Ti:sapphire gain is low. The primary condition for broadband spectrum is attributed to operating this compact laser cavity at small negative net GDD. However precise oscillation compensation has not been important and the net cavity GDD is found to be positive near the Ti:sapphire gain peak. The use of a slightly convex mirror contributed to additional broadening. We expect that this ultra-broadband laser with improved IR spectrum will make easier to measure and stabilize the laser carrier-to-envelope offset frequency by use of nonlinear f-2f spectrometers[1], without use of microstructure photonic fibers.

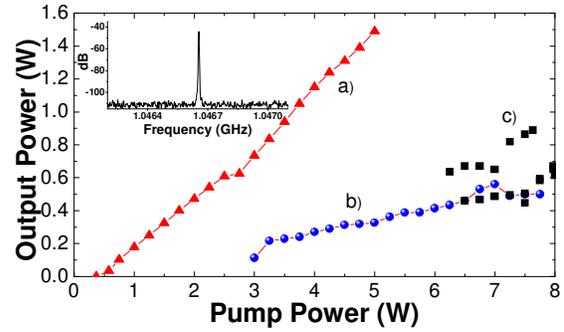

Fig. 3. Laser output powers for the 4-mirror bow-tie cavity. a) Red (online) triangles: cw operation; blue (online) circles: mode-locked operation in the first region discussed in the text, corresponding to curve b of Fig.1; black (online) squares: mode-locked operation in the second region discussed in the text (curve d in Fig.1). Inset: beatnote at laser repetition rate.

The authors thank Scott Diddams for valuable contributions, including discussions and equipment loan; Tara Fortier, Matt Kirchner and Michael Thorpe for contributions on the setup, measurements and analysis software for WLI; Frank Wunderlich (Layertec) for providing data on mirrors; and Seth Foreman and Henry Kaptein for comments on the manuscript. FCC (flavio@ifi.unicamp.br) is a 2005 JILA visiting fellow and he also acknowledges the support of FAPESP, CAPES and CNPq - Brazil. GTN acknowledges a scholarship from CAPES - Brazil and the support of NIST during a four months stay in Boulder.

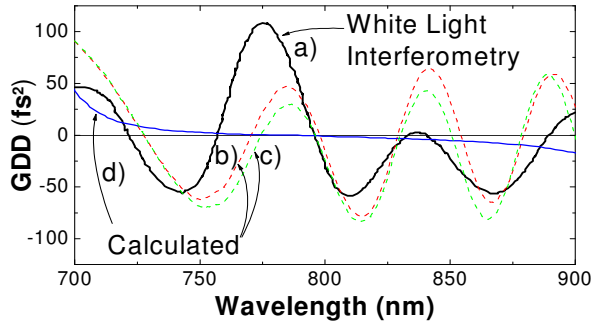

Fig.1. Net group delay dispersion (GDD) of the 4-mirror bow tie laser cavity including the crystal and excluding the output coupler. a) Black (online) solid: measurement by white-light interferometry, with convex mirror included; b) Red (online) dashed: calculated GDD of mirrors based on curves provided by manufacturer (convex mirror included), combined with a Ti:sapphire 3-mm crystal flat GDD of 174 fs$^2$ [12]; c) green (online) dashed: the same as b), but with convex mirror replaced by HR flat mirror; d) calculated GDD of the OC provided by the manufacturer.

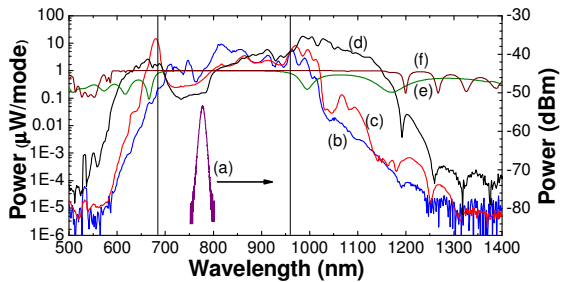

Fig. 2 Output spectra for the Ti:sapphire mode-locked lasers. a) Purple (online) solid: normal mode-locked laser using the 6-mirror ring cavities with excess negative GDD (see text); b) Blue (online) solid: broad spectrum obtained with the 4-mirror cavity with small net GDD; c) and d) red and black (online) solid: spectra when a convex mirror (R=1 m) replaces the HR flat mirror, for two different alignment conditions; e) and f) olive and brown (online): reflectivity of the OC and IC (read using the numbers on left Y axis). Vertical bars indicate the laser cavity bandwidth, defined at the points where the OC transmission increases to 20%.